\documentclass[a4,11pt]{article}
\usepackage{latexsym,enumerate}
\usepackage{amsmath,amsthm,amsopn,amstext,amscd,amsfonts,amssymb}
\usepackage{natbib,graphicx,fullpage,subfigure}
\usepackage{hyperref}
\usepackage{amssymb}
\usepackage{graphicx}
\usepackage{bbm}

\begin{document}

\noindent
{\Large \bf Discussion on ``Sparse graphs using exchangeable random measures" by F.~Caron and E.B.~Fox, written by Roberto Casarin$^{\dag}$, Matteo Iacopini$^{\dag \S}$\footnote[1]{Corresponding author at: Ca' Foscari University of Venice, Cannaregio 873, 30121, Venice, Italy. \\ \textit{E-mail address}: \href{matteo.iacopini@unive.it}{matteo.iacopini@unive.it} (Matteo Iacopini)} and Luca Rossini$^{\dag \ddag}$({\large $^\dag$ University Ca' Foscari of Venice, $^\S$ Universit\'{e} Paris 1 - Panth\'{e}on-Sorbonne and $^\ddag$ Free University of Bozen-Bolzano}).}

\vspace{10pt}

The authors are to be congratulated on their excellent research, which has culminated in the development of a new class of random graph models. The node degree and the degree distribution fail in giving a unique characterisation of network complexity (\cite{PhysRevE.82.066102}). For this reason global connectivity measures, such as communicability \citep{Est1,Est2} and centrality (\cite{Borgatti2006466}) are used to analyse a graph. In this discussion we contribute to the analysis of the GGP model as compared to the Erd{\"o}s-Renyi (ER) and the preferential attachment (AB) (\cite{Barabasi99}) models. Our analysis is far from being exhaustive, but shows that more theoretical aspects of the GGP model are to be investigated.

A connected component of the $n$-nodes graph $G=(V,E)$ is a subgraph in which any two vertices $v_i$ and $v_j$ are connected by paths. The number of connected components equals the multiplicity of the null eigenvalue of the graph Laplacian $L$, where the $(i,j)$ entry of $L$ is:
\begin{eqnarray*}
L_{ij}:=\left\lbrace\begin{array}{ccc}
d(v_i) && \mbox{if } i=j , \\ -1 && \mbox{if } i \ne j \mbox{ and } (v_i,v_j)\in E, \\ 0 && \mathrm{otherwise},
\end{array}\right.
\end{eqnarray*}
with $d(v_i)$ the degree of $v_i$.

The global clustering coefficient measures the tendency of nodes to cluster together and is defined as:
\begin{equation}
C = \frac{\mbox{number of triangle loops}}{\mbox{number of connected triples of vertices}}. \notag
\end{equation}

The assortativity coefficient between pairs of linked nodes is given by:
\begin{equation*}
r = \frac{\sum_{j=1}^{n}\sum_{k=1}^{n} jk(e_{jk}-q_j q_k)}{\sigma^2_q},
\end{equation*}
where $q_k, \, e_{jk}$ are the distribution and the joint-excess degree probability of the remaining degrees, respectively, for the two vertices $v_j$ and $v_k$ and $\sigma_q$ is the standard deviation of $q_k$.

Finally, given the partition of the network into two non-overlapping subgraphs (core and periphery) that maximizes the number/weight of within core-group edges, we compute the share of nodes in the core.

\begin{figure}[t]
\centering
\begin{tabular}{cc}
\includegraphics[width=6.85cm]{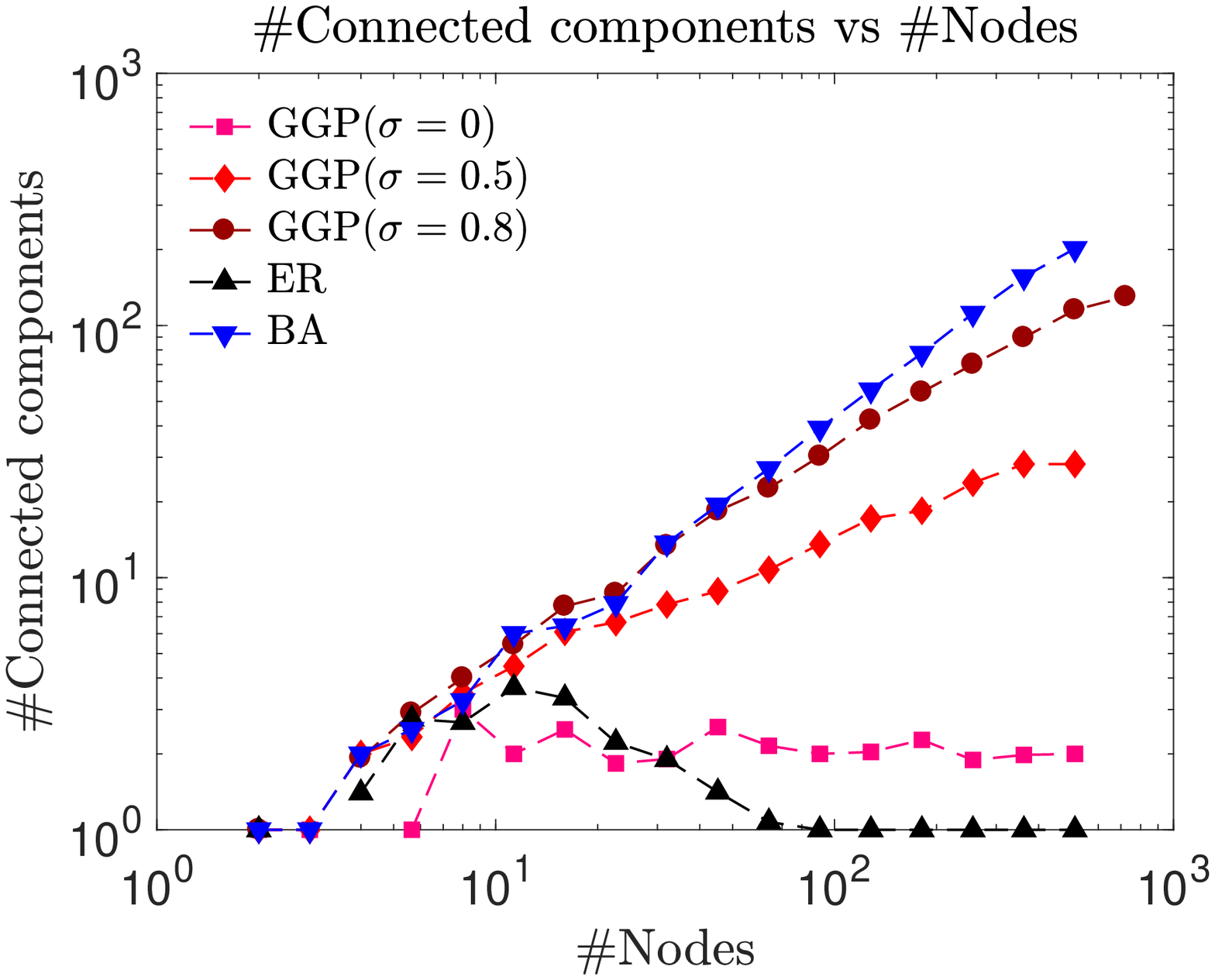}&
\includegraphics[width=6.85cm]{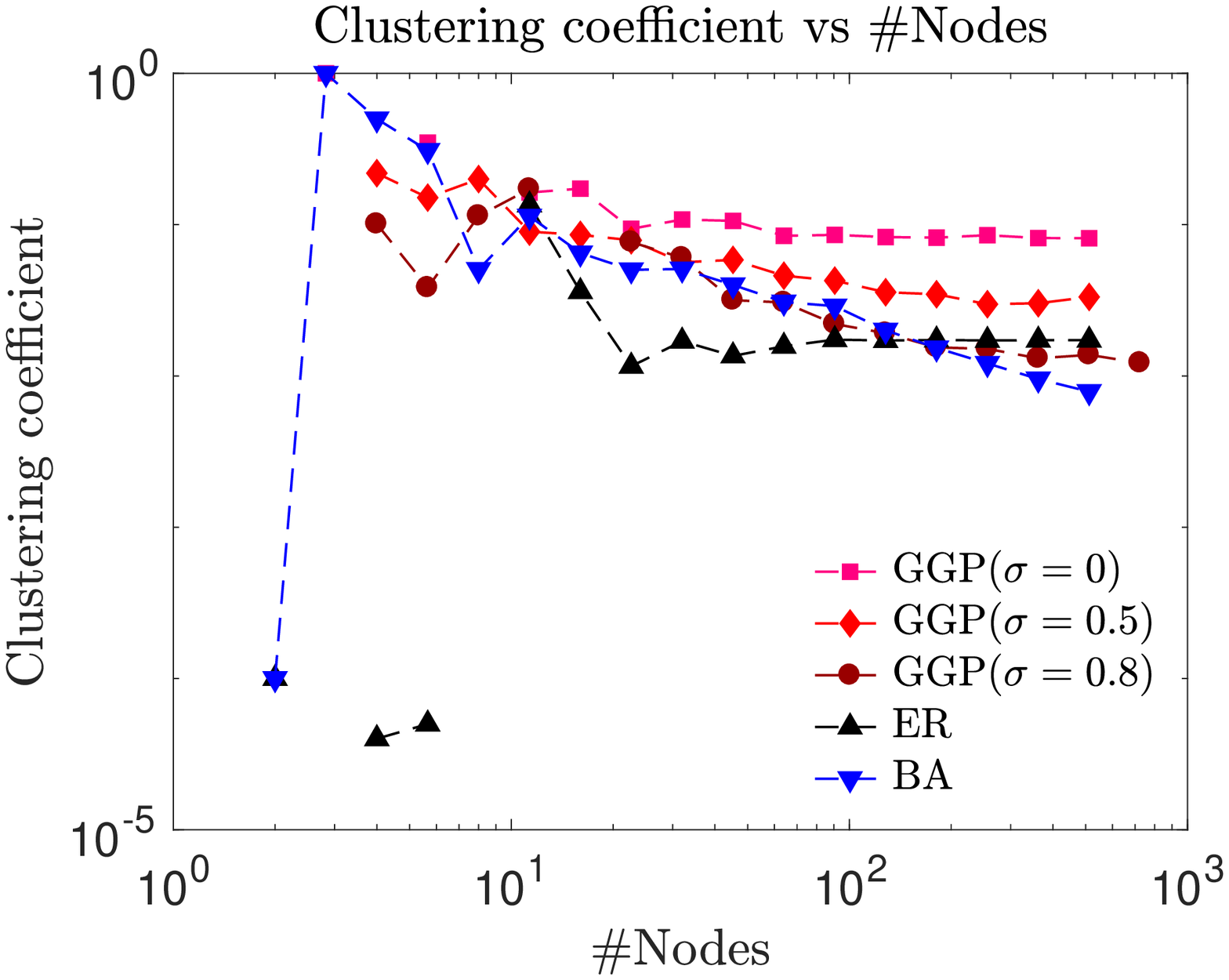}\\
(a)&(b)\\
\includegraphics[width=6.85cm]{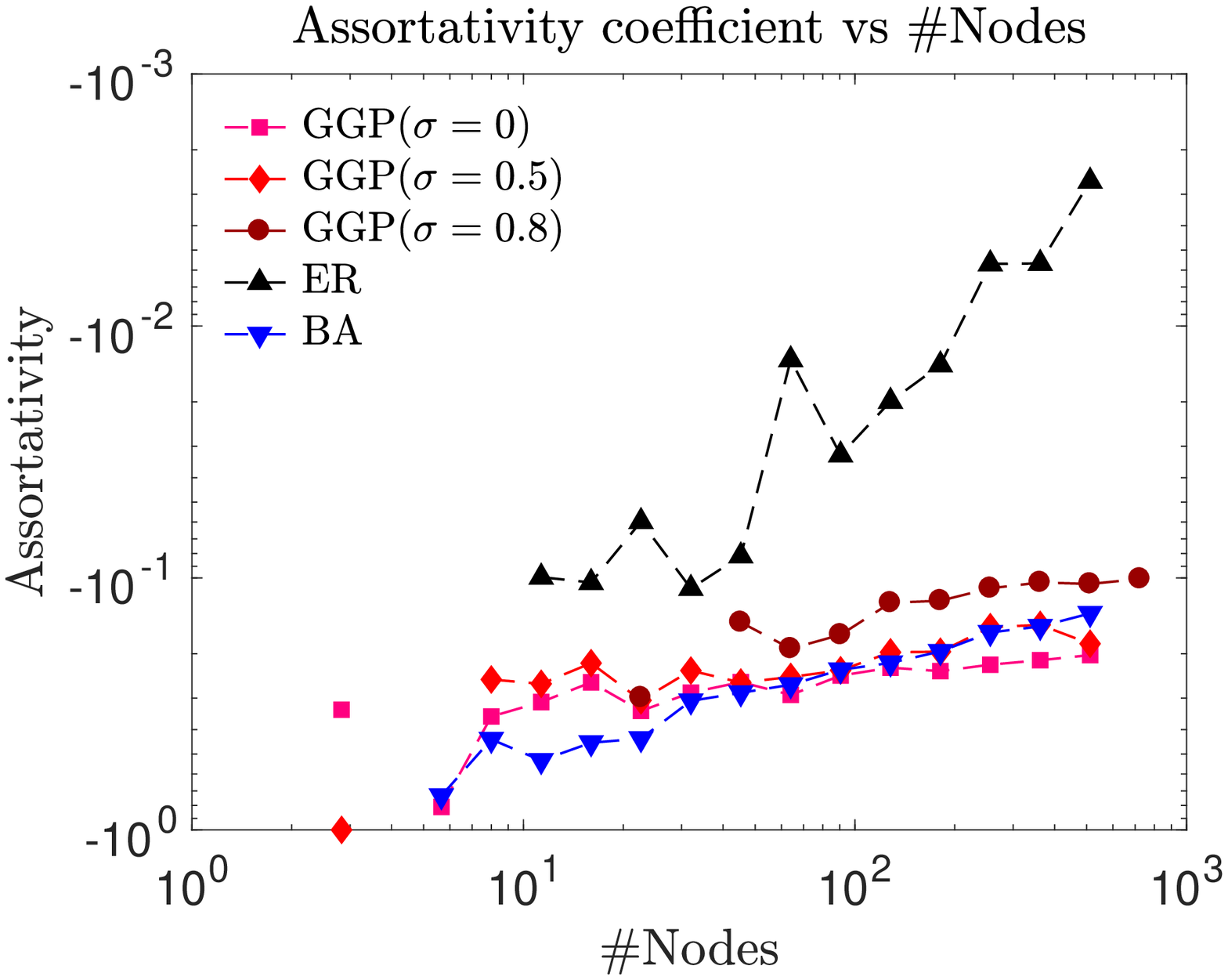}&\includegraphics[width=6.85cm]{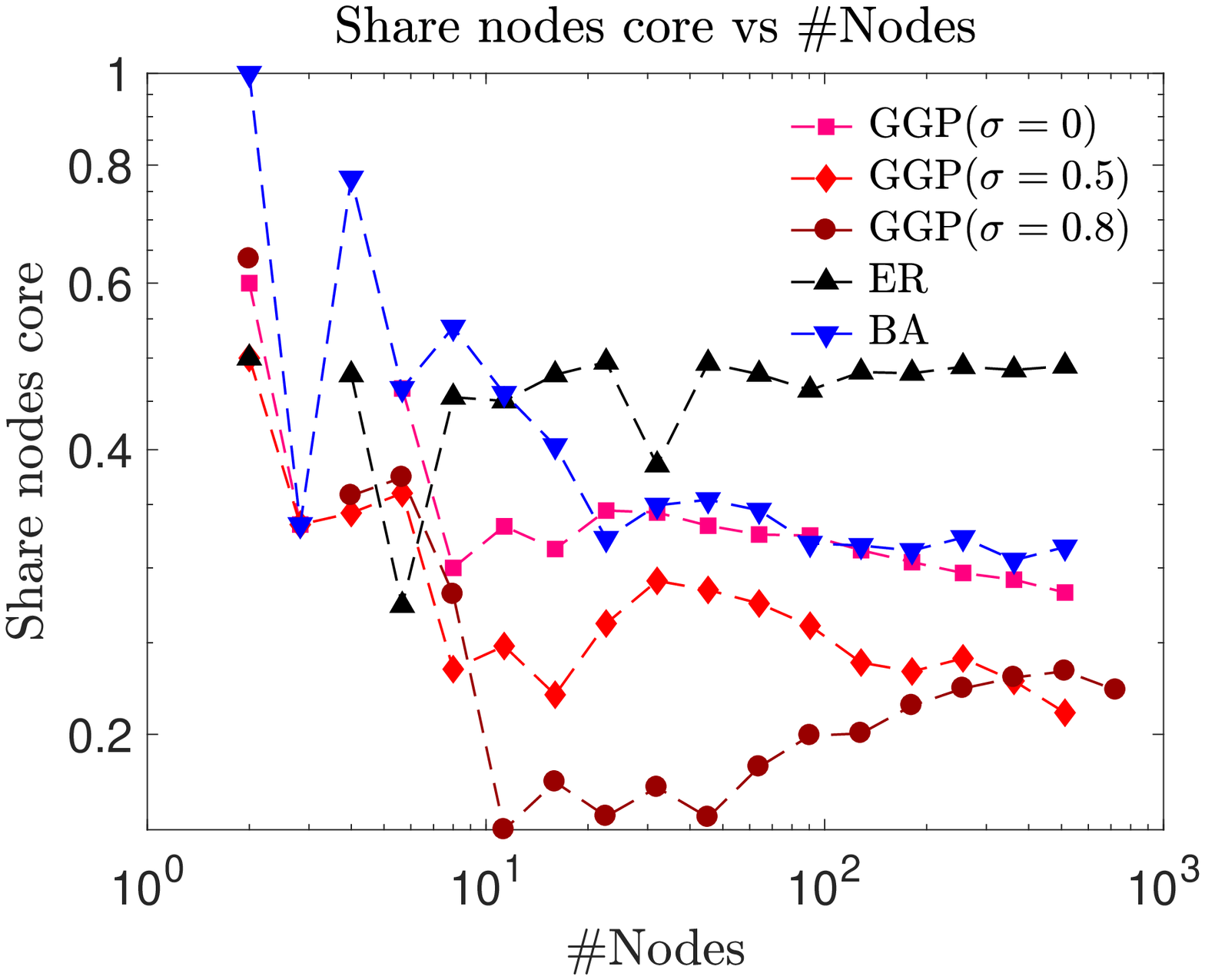}\\
(c)&(d)\\
\end{tabular}
\caption{Network statistics versus number of nodes for the GGP undirected network, the Erd{\"o}s-Renyi (ER) and the preferential attachment model of \cite{Barabasi99} (AB). Panel (a): number of connected components; Panel (b): clustering coefficient; Panel (c): assortativity coefficient; Panel (d): share nodes core.}\label{Fig}
\end{figure}

According to Figure \ref{Fig} panel (a), the GGP couples with the AB model and performs slightly worse with the ER random graph in terms of the number of connected components. Panels (b) and (c) highlight that the clustering structure of GGP does not vary too much with $\sigma$. The clustering coefficient is in line with the two benchmarks while the assortativity of the ER model is not attained. For $\sigma=0.5,0.8$, GGP exhibits a lower share of nodes in the core (panel (d)) than in the benchmarks, and mimics the AB model for $\sigma=0$.

Overall, the GGP can replicate typical behaviours of real world sparse networks and some fundamental features of random graphs generated from the AB model, making it suitable for a variety of applications in different fields.

We are very pleased to be able to propose the vote of thanks to the authors for their work.

\bibliographystyle{apalike}
\bibliography{Refer}

\begin{thebibliography}{}

\bibitem[Barabasi and Albert, 1999]{Barabasi99}
Barabasi, A.~L. and Albert, R. (1999).
\newblock Emergence of scaling in random networks.
\newblock {\em Science}, 286(5439):509--512.

\bibitem[Borgatti and Everett, 2006]{Borgatti2006466}
Borgatti, S.~P. and Everett, M.~G. (2006).
\newblock A graph-theoretic perspective on centrality.
\newblock {\em Social Networks}, 28(4):466--484.

\bibitem[Estrada, 2010]{PhysRevE.82.066102}
Estrada, E. (2010).
\newblock Quantifying network heterogeneity.
\newblock {\em Physical Review E}, 82(6):066102--8.

\bibitem[Estrada and Hatano, 2008]{Est1}
Estrada, E. and Hatano, N. (2008).
\newblock Communicability in complex networks.
\newblock {\em Physical Review E}, 77(3):036111--12.

\bibitem[Estrada and Hatano, 2009]{Est2}
Estrada, E. and Hatano, N. (2009).
\newblock Communicability graph and community structures in complex networks.
\newblock {\em Applied Mathematics and Computation}, 214(2):500--511.

\end{thebibliography}

%
%
%
%
%


\end{document}